\documentclass{article}
\usepackage{amsmath}
\usepackage[psamsfonts]{amssymb}
\usepackage{cmmib57}
\usepackage{diagrams}
\newcommand{\bPf}{\par\vspace*{-4pt}\indent{\sc Proof.}\enskip}
\newcommand{\ePf}{\medskip}
\def\QED{\hskip0.1em\hfill\null\ \null\nobreak\hfill\kern3pt\vbox{\hrule\hbox
   {\vrule\kern1pt\vbox{\kern1.7pt\hbox{$\scriptscriptstyle{QED}$}
    \kern0.2pt}\kern1pt\vrule}\hrule}}

\def\END{\hskip0.1em\hfill\null\ \null\nobreak\hfill\kern3pt\vbox{\hrule\hbox
   {\vrule\kern1pt\vbox{\kern1.7pt\hbox{$\,\,\,\vspace{5pt}$}
    \kern0.2pt}\kern1pt\vrule}\hrule}}
\newtheorem{theorem}{Theorem}
\newtheorem{lemma}{Lemma}
\newtheorem{corollary}{Corollary}
\newtheorem{proposition}{Proposition}
\newtheorem{remark}{Remark}
\newtheorem{definition}{Definition}
\newtheorem{example}{Example}

\newcommand{\olin}[1]{\overline{#1}}

\newcommand{\bCd}{\bEq\begin{CD}}
\newcommand{\eCd}{\end{CD}\eEq}
\newcommand{\bcd}{\beq\begin{CD}}
\newcommand{\ecd}{\end{CD}\eeq}
\newcommand{\ben}{\begin{enumerate}}
\newcommand{\een}{\end{enumerate}}
\newcommand{\bEq}{\begin{eqnarray}}
\newcommand{\eEq}{\end{eqnarray}}
\newcommand{\beq}{\begin{eqnarray*}}
\newcommand{\eeq}{\end{eqnarray*}}
\newcommand{\bDf}{\begin{definition}\em}
\newcommand{\eDf}{\end{definition}}
\newcommand{\bLm}{\begin{lemma}}
\newcommand{\eLm}{\end{lemma}}
\newcommand{\bPr}{\begin{proposition}}
\newcommand{\ePr}{\end{proposition}}
\newcommand{\bTh}{\begin{theorem}}
\newcommand{\eTh}{\end{theorem}}
\newcommand{\bCr}{\begin{corollary}}
\newcommand{\eCr}{\end{corollary}}
\newcommand{\bRm}{\begin{remark}\em}
\newcommand{\eRm}{\end{remark}}
\newcommand{\bEx}{\begin{example}\em}
\newcommand{\eEx}{\end{example}}

\newcommand{\ie}{{\em i.e$.$} }

\newcommand{\R}{I\!\!R}

\newcommand{\mto}{\mapsto}

\newcommand{\der}{\partial}

\DeclareMathOperator{\im}{im}
\DeclareMathOperator{\byd}{{\raisebox{.1ex}{:}{=}}}

\newcommand{\ucar}[1]{\underset{#1}{\times}}

\newcommand{\owed}[1]{\overset{#1}{\wedge}}

\newcommand{\balp}{\boldsymbol{\alp}}
\newcommand{\bbet}{\boldsymbol{\bet}}
\newcommand{\bgam}{\boldsymbol{\gam}}

\newcommand{\blam}{\boldsymbol{\lam}}

\newcommand{\bnu}{\boldsymbol{\nu}}

\newcommand{\cA}{\mathcal{A}}

\newcommand{\cC}{\mathcal{C}}
\newcommand{\cD}{\mathcal{D}}
\newcommand{\cE}{\mathcal{E}}

\newcommand{\cL}{\mathcal{L}}

\newcommand{\cN}{\mathcal{N}}

\newcommand{\cT}{\mathcal{T}}

\newcommand{\cZ}{\mathcal{Z}}

\newcommand{\bx}{\boldsymbol{x}}
\newcommand{\by}{\boldsymbol{y}}

\newcommand{\bA}{\boldsymbol{A}}

\newcommand{\bC}{\boldsymbol{C}}

\newcommand{\bF}{\boldsymbol{F}}
\newcommand{\bG}{\boldsymbol{G}}

\newcommand{\bP}{\boldsymbol{P}}

\newcommand{\bU}{\boldsymbol{U}}

\newcommand{\bW}{\boldsymbol{W}}
\newcommand{\bX}{\boldsymbol{X}}
\newcommand{\bY}{\boldsymbol{Y}}

\newcommand{\sub}{\subset}

\newcommand{\wed}{\wedge}
\newcommand{\com}{\!\circ\!}
\newcommand{\ten}{\!\otimes\!}

\newcommand{\alp}{\alpha}
\newcommand{\bet}{\beta}
\newcommand{\gam}{\gamma}
\newcommand{\del}{\delta}
\newcommand{\eps}{\epsilon}
\newcommand{\zet}{\zeta}

\newcommand{\lam}{\lambda}
\newcommand{\sig}{\sigma}

\newcommand{\ome}{\omega}
\newcommand{\Gam}{\Gamma}

\newcommand{\Lam}{\Lambda}

\newcommand{\vartht}{\vartheta}

\newcommand{\For}{{\Lambda}}
\newcommand{\Con}{{\mathcal{C}}}
\newcommand{\Hor}{{\mathcal{H}}}
\newcommand{\Var}{{\mathcal{V}}}
\newcommand{\Thd}{{\Theta}}
\title{\large{{\bf Conservation Laws and Variational 
Sequences in Gauge--Natural Theories}}\thanks{Work partially supported by GNFM 
of INDAM, MURST, University of 
Turin.}}
\author{{\normalsize L. Fatibene, M. Francaviglia and M. Palese}
\\{\footnotesize Department of Mathematics,
University of Turin}
\\{\footnotesize Via C. Alberto 10, 10123 Turin, Italy}\\ 
{\footnotesize e--mails: 
{\sc fatibene@dm.unito.it, francaviglia@dm.unito.it,}}\\
{\footnotesize {\sc palese@dm.unito.it}}}
\date{}
\begin{document}

\maketitle

\begin{abstract}
\footnotesize{In the classical Lagrangian approach 
to conservation laws of gauge-natural field theories
a suitable (vector) density is known to generate the so--called 
{\em conserved Noether currents}. 
It turns out that along any section of the relevant gauge--natural bundle 
this density is the divergence of a skew--symmetric (tensor) density, 
which is called a {\em superpotential} for the conserved currents.

We describe gauge--natural superpotentials in the 
framework of finite order variational sequences according to Krupka.
We refer to previous results of ours on {\em variational Lie derivatives} 
concerning abstract versions of Noether's theorems, 
which are here interpreted in terms of ``horizontal'' and 
``vertical'' conserved currents.  
The gauge--natural lift of principal automorphisms 
implies suitable linearity properties of the Lie derivative operator. 
Thus abstract results due to Kol\'a\v{r}, concerning the
integration by parts procedure, can be applied to prove the
{\em existence} and {\em globality} of superpotentials in a very general setting. 

\medskip


}
\end{abstract}

\small

\section{Introduction}

Our framework is the calculus of variations on finite order fibered 
manifolds. In the classical Lagrangian 
formulation of field theories the description of symmetries 
amounts to define a suitable vector density 
which generates the conserved currents; 
this density is found to be the divergence of a skew--symmetric tensor density, 
which is called a {\em superpotential} for the conserved currents 
(see {\em e.g.} \cite{Fa99,FeFr91} and references quoted therein).

It is well known that symmetries enable us to better understand the structure 
of physical theories. It is also known that in general physical fields 
are not natural objects, \ie there is no canonical way of 
associating a transformation of the relevant
configuration bundle to each diffeomorphism of its basis
(conventionally ``space--time''). In order to 
apply Noether's theorems to provide conserved quantities 
(like energy, momentum and 
angular momentum) for these theories we thence need further structures 
to provide an operative definition of the notion of 
``horizontal'' symmetries. On the other hand, ``vertical'' symmetries are 
associated to others conserved quantities, like {\em e.g.} 
``charges'' in gauge theories. 

The gauge--natural setting \cite{Ec81,Fa99,Ja90,KMS93} is the most suitable 
one for an exhaustive description of this physical situation.
{\em Gauge--natural bundles} of any given order, 
turn out to be fiber bundles associated to a suitable prolongation of a
principal bundle $\bP$ with an arbitrary structure group $\bG$
\cite{Ec81,KMS93}. The ``structure bundle'' $\bP$ encodes global properties of
the group of so--called gauge transformations so that a canonical treatment of
gauge symmetries  can be achieved. In previously existing frameworks
symmetries were, in fact, defined as forming an arbitrary (often unspecified) subgroup 
of automorphisms of the associated bundle. On the contrary, 
in a gauge--natural theory the existence of preferred gauge--natural lifts of 
automorphisms of $\bP$ implies that the very knowledge of the structure bundle $\bP$
of the associated bundle 
allows a suitable unifying description of gauge invariance and covariance 
properties, thence of conserved quantities.

The subgroup of so--called {\em pure gauge transformations}
corresponds to vertical automorphisms of the structure bundle $\bP$.
However, there is no canonical way to define a {\em complementary} subgroup 
of {\em horizontal} automorphisms. Dynamical connections over the 
structure bundle (\ie connections over the configuration bundle induced 
by principal connections over $\bP$ together with linear 
connections over the frame bundle $L(\bX)$) enable us to define the 
{\em infinitesimal horizontal transformations} which are related 
to conserved quantities such as energy, momentum and angular momentum.
Dynamical connections can also be used to solve some ambiguities in 
the definition of the Poincar\'e--Cartan 
morphism, conserved currents and superpotentials \cite{FPV98b,Vit98}.

Following \cite{Kru90,Kru93}, we consider one of the recent geometrical
formulations of the Calculus of  Variations on finite order jets of fibered
manifolds. As it is well known, in this formulation
the {\em  variational sequence\/} is defined as a quotient of the de 
Rham sequence on a finite order jet space  with respect to an 
intrinsically defined subsequence, the ``contact'' subsequence, so 
that only ``variationally relevant'' objects are defined there (see 
{\em e.g.} \cite{Pa00,Vit97,Vit98}).

In a previous work \cite{FPV98b} we already provided a
description of superpotentials in the framework of 
variational sequences for natural field theories. In this paper we shall 
generalize these 
results to the much larger class of gauge--natural theories, showing 
how the (non--natural) added structure reflects into the framework. 
We make use of the representation of the quotient 
sheaves of the variational sequence as concrete sheaves of forms
given in \cite{Vit98}
and of previous results of ours on {\em variational Lie derivatives} 
which provide suitable formulations of Noether's Theorems \cite{FPV98a}.
Furthermore, we refer to some abstract results concerning {\em 
global} decomposition formulae of morphisms, involved with 
the integration by parts procedure \cite{HoKo83,Kol83,Kol84}. In 
order to apply this results, we stress linearity properties of 
the Lie derivative operator, which rely on properties 
of the gauge--natural lift of principal automorphisms. In this way
we provide an intrinsic proof of {\em existence} and {\em globality} 
of superpotentials for gauge--natural theories.

\medskip

Here, manifolds and maps between manifolds are $C^{\infty}$.
All morphisms of fibered manifolds (and hence bundles) will 
be morphisms over the identity of the base manifold, unless 
otherwise specified. As for sheaves, we will use the definitions 
and the main results given in \cite{Wel80}.

\section{Jet spaces and variational sequences}\label{2}

In this Section we recall some basic facts about jet spaces
\cite{Fe84,MaMo83a,Sau89} and Krupka's formulation of the finite
order variational sequence \cite{Kru90,Vit98}.

\subsection{Jet spaces}\label{notations}

Here we introduce jet spaces of a
fibered manifold and the sheaves of forms on the $r$--th order
jet space. Moreover, we recall the notion of horizontal and vertical
differential \cite{Sau89}.

Our framework is a fibered manifold $\pi : \bY \to \bX$,
with $\dim \bX = n$ and $\dim \bY = n+m$.

For $r \geq 0$ we are concerned with the $r$--jet space $J_r\bY$;
in particular, we set $J_0\bY \equiv \bY$. We recall the natural fiberings
$\pi^r_s : J_r\bY \to J_s\bY$, $r \geq s$, $\pi^r : J_r\bY \to \bX$, and,
among these, the {\em affine\/} fiberings $\pi^r_{r-1}$.
We denote with $V\bY$ the vector subbundle of the tangent
bundle $T\bY$ of vectors on $\bY$ which are vertical with respect
to the fibering $\pi$.

Charts on $\bY$ adapted to $\pi$ are denoted by $(x^\sig ,y^i)$.  Greek
indices $\sig ,\mu ,\dots$ run from $1$ to $n$ and they label base
coordinates, while
Latin indices $i,j,\dots$ run from $1$ to $m$ and label fibre coordinates,
unless otherwise specified.  We denote by $(\der_\sig ,\der_i)$ and 
$(d^\sig, d^i)$ the local bases of vector fields and $1$--forms on $\bY$
induced by an adapted chart, respectively.

We denote multi--indices of dimension $n$ by boldface Greek letters such as
$\balp = (\alp_1, \dots, \alp_n)$, with $0 \leq \alp_\mu$,
$\mu=1,\ldots,n$; by an abuse
of notation, we denote with $\sig$ the multi--index such that
$\alp_{\mu}=0$, if $\mu\neq \sig$, $\alp_{\mu}= 1$, if
$\mu=\sig$.
We also set $|\balp| \byd \alp_{1} + \dots + \alp_{n}$ and $\balp ! \byd
\alp_{1}! \dots \alp_{n}!$.

The charts induced on $J_r\bY$ are denoted by $(x^\sig,y^i_{\balp})$, with $0
\leq |\balp| \leq r$; in particular, we set $y^i_{\bf{0}}
\equiv y^i$. The local vector fields and forms of $J_r\bY$ induced by
the above coordinates are denoted by $(\der^{\balp}_i)$ and $(d^i_{\balp})$,
respectively.

In the theory of variational sequences a fundamental role is played by the
{\em contact maps\/} on jet spaces (see \cite{Fe84,Kru90,Kru93,MaMo83a}).
Namely, for $r\geq 1$, we consider the natural complementary fibered
morphisms over $J_r\bY \to J_{r-1}\bY$
\beq
\cD : J_r\bY \ucar{\bX} T\bX \to TJ_{r-1}\bY \,,
\qquad
\vartht : J_r\bY \ucar{J_{r-1}\bY} TJ_{r-1}\bY \to VJ_{r-1}\bY \,,
\eeq
with coordinate expressions, for $0 \leq |\balp| \leq r-1$, given by
\beq
\cD &= d^\lam\ten {\cD}_\lam = d^\lam\ten
(\der_\lam + y^j_{\balp+\lam}\der_j^{\balp}) \,,
\vartht &= \vartht^j_{\balp}\ten\der_j^{\balp} =
(d^j_{\balp}-y^j_{{\balp}+\lam}d^\lam)
\ten\der_j^{\balp} \,.
\eeq

We have 
\bEq
\label{jet connection}
J_r\bY\ucar{J_{r-1}\bY}T^*J_{r-1}\bY =\left(
J_r\bY\ucar{J_{r-1}\bY}T^*\bX\right) \oplus\cC^{*}_{r-1}[\bY]\,,
\eEq
where $\cC^{*}_{r-1}[\bY] \byd \im \vartht_r^*$ and
$\vartht_r^* : J_r\bY \ucar{J_{r-1}\bY} V^*J_{r-1}\bY \to
J_r\bY \ucar{J_{r-1}\bY} T^*J_{r-1}\bY \,$. We have the isomorphism 
$\cC^{*}_{r-1}[\bY] \simeq J_{r}\bY \ucar{J_{r-1}\bY} V^{*}J_{r-1}\bY$.

If $f: J_{r}\bY \to \R$ is a function, then we set
$D_{\sig}f$ $\byd \cD_{\sig} f$,
$D_{\balp+\sig}f$ $\byd D_{\sig} D_{\balp}f$, where $D_{\sig}$ is
the standard {\em formal derivative}.
Given a vector field $Z : J_{r}\bY \to TJ_{r}\bY$, the splitting
\eqref{jet connection} yields $Z \com \pi^{r+1}_{r} = Z_{H} + Z_{V}$
where, if $Z = Z^{\gam}\der_{\gam} + Z^i_{\balp}\der^{\balp}_i$, then we
have $Z_{H} = Z^{\gam}D_{\gam}$ and
$Z_{V} = (Z^i_{\balp} - y^i_{\balp + \gam}Z^{\gam}) \der^{\balp}_{i}$.

The splitting
\eqref{jet connection} induces also a decomposition of the
exterior differential on $\bY$,
$(\pi^{r+1}_r)^*\com \,d = d_H + d_V$, where $d_H$ and $d_V$
are defined to be the {\em horizontal\/} and {\em vertical differential\/}.
The action of $d_H$ and $d_V$ on functions and $1$--forms
on $J_r\bY$ uniquely characterizes $d_H$ and $d_V$ (see, {\em e.g.},
\cite{Sau89,Vit98} for more details).

A {\em projectable vector field\/} on $\bY$ is defined to be a pair
$(\Xi,\xi)$, where $\Xi:\bY \to T\bY$ and $\xi: \bX \to T\bX$
are vector fields and $\Xi$ is a fibered morphism over $\xi$.
If there is no danger of confusion, we will denote simply by $\Xi$ a
projectable vector field $(\Xi,\xi)$.

A projectable vector field $(\Xi,\xi)$, with coordinate expression
$\Xi = \xi^{\sig}\der_{\sig} + \xi^i\der_{i}$,
$\xi = \xi^{\sig}\der_{\sig}$,
can be conveniently prolonged to a projectable vector field
$(j_{r}\Xi, \xi)$, whose coordinate expression turns out to be
\beq
j_{r}\Xi = \xi^{\sig}\der_{\sig} +
\left(
D_{\balp}\xi^i - \sum_{\bbet + \bgam = \balp}
\frac{\balp !}{\bbet ! \bgam !} D_{\bbet}\xi^{\mu} \, y^i_{\bgam + \mu}
\right) \, \der_i^{\balp} \,,
\eeq
where $\bbet \neq 0$ and $0 \leq |\balp | \leq r$ (see
\cite{MaMo83a,Sau89});
in particular, we have the following expressions
$(j_{r}\Xi)_{H} = \xi^{\sig} \, D_{\sig}$,
$(j_{r}\Xi)_{V} = D_{\balp}(\Xi_{V})^i \, \der_i^{\balp}$,
with $(\Xi_{V})^i = \xi^i - \, y^i_{\sig}\xi^{\sig}$.
From now on, by an abuse of notation, we will write simply $j_{r}\Xi_{H}$ and
$j_{r}\Xi_{V}$.

\subsection{Variational sequences}

We shall be here concerned with some distinguished sheaves of forms on jet
spaces \cite{Fe84,Kru90,Kru93,Sau89,Vit98}.
Due to the topological triviality of the fibre of
$J_{r}\bY\to \bY$, we will consider sheaves on $J_{r}\bY$ with respect to
the topology generated by open sets of the kind
$\left( {\pi_0^r}\right)^{-1}(\bU)$, with $\bU\subset\bY$ open in $\bY$.

i. For $r \geq 0$, we consider the standard sheaves $\For^{p}_r$
of $p$--forms on $J_r\bY$.

ii. For $0 \leq s \leq r $, we consider the sheaves $\Hor^{p}_{(r,s)}$ and
$\Hor^{p}_r$ of {\em horizontal forms\/}, \ie of local fibered morphisms over
$\pi^{r}_{s}$ and $\pi^{r}$ of the type
$\alp : J_r\bY \to \owed{p}T^*J_s\bY$ and $\bet : J_r\bY \to \owed{p}T^*\bX$,
respectively.

iii. For $0 \leq s < r$, we consider the subsheaf $\Con^{p}_{(r,s)}
\sub \Hor^{p}_{(r,s)}$ of {\em contact forms\/}, \ie
of sections $\alp \in \Hor^{p}_{(r,s)}$ with values into
$\owed{p} (\im\vartht_{s+1}^*)$.
We have the distinguished subsheaf $\Con^{p}{_r} \sub
\Con^{p}_{(r+1,r)}$ of local fibered morphisms $\alp \in \Con^{p}_{(r+1,r)}$
such that $\alp = \owed{p}\vartht_{r+1}^* \com \Tilde{\alp}$, where
$\tilde{\alp}$ is a section of the fibration $J_{r+1}\bY \ucar{J_r\bY}$
$\owed{p}V^*J_r\bY$ $\to J_{r+1}\bY$ which projects down onto
$J_{r}\bY$.

According to \cite{Vit98}, the fibered splitting
\eqref{jet connection} yields the sheaf splitting
$\Hor^{p}_{(r+1,r)}$ $=$ $\bigoplus_{t=0}^p$
$\Con^{p-t}_{(r+1,r)}$ $\wed\Hor^{t}_{r+1}$, which restricts to the inclusion
$\For^{p}_r$ $\sub$ $\bigoplus_{t=0}^{p}$
$\Con^{p-t}{_r}\wed\Hor^{t}{_{r+1}^{h}}$,
where $\Hor^{p}{_{r+1}^{h}}$ $\byd$ $h(\For^{p}_r)$ for $0 < p\leq n$ and
$h$ is defined to be the restriction to $\For^{p}_{r}$ of the projection of
the above splitting onto the non--trivial summand with the highest
value of $t$. We define also the map $v \byd \textstyle{id} - h$.

We recall now the theory of variational sequences on finite order jet
spaces, as it was developed by Krupka in \cite{Kru90}.

By an abuse of notation, let us denote by $d\ker h$ the sheaf
generated by the presheaf $d\ker h$ (see \cite{Wel80}).
We set $\Thd^{*}_{r}$ $\byd$ $\ker h$ $+$
$d\ker h$.

In \cite{Kru90} it was proved that the following diagram is 
commutative and that its rows and columns are exact:
\beq
\diagramstyle[size=1.3em]
\begin{diagram}
&& 0 && 0 && 0 && 0 &&&& 0 && 0  &&&&
\\
&& \dTo && \dTo && \dTo && \dTo &&&& \dTo && \dTo &&&&
\\
0 & \rTo & 0 & \rTo & 0 & \rTo &
\Thd^{1}_r & \rTo^d & \Thd^{2}_r & \rTo^d & \dots &
\rTo^d & \Thd^{I}_r & \rTo^d & 0 & \rTo & \dots & \rTo & 0
\\
&& \dTo && \dTo && \dTo && \dTo &&&& \dTo && \dTo &&&&
\\
0 & \rTo & \R & \rTo & \For^{0}_r & \rTo^d &
\For^{1}_r & \rTo^d & \For^{2}_r & \rTo^d & \dots & \rTo^d &
\For^{I}_r & \rTo^d & \For^{I+1}_r & \rTo^d & \dots & \rTo^d & 0
\\
&& \dTo && \dTo && \dTo && \dTo &&&& \dTo && \dTo &&&&
\\
0 & \rTo & \R & \rTo & \For^{0}_r & \rTo^{\cE_{0}} &
\For^{1}_r/\Thd^{1}_r & \rTo^{\cE_{1}} & \For^{2}_r/\Thd^{2}_r & \rTo^{\cE_{2}} &
\dots & \rTo^{\cE_{I-1}} & \For^{I}_r/\Thd^{I}_r & \rTo^{\cE_{I}} &
\For^{I+1}_r & \rTo^{d} & \dots & \rTo^{d} & 0
\\
&& \dTo && \dTo && \dTo && \dTo &&&& \dTo && \dTo &&&&
\\
&& 0 && 0 && 0 && 0 &&&& 0 && 0  &&&&
\end{diagram}
\eeq

\bDf
The top row of the above diagram is said to be the {\em contact sequence};
the bottom row is said to be the $r$--th order
{\em variational sequence\/} associated with the fibered manifold
$\bY\to\bX$.
\END
\eDf

Notice that, in general, the highest integer $I$ depends on the dimension
of the fibers of $J_{r}\bY \to \bX$.

We can consider the `short' variational sequence: 
\beq
\diagramstyle[size=1.3em]
\begin{diagram}
0 &\rTo & \R &\rTo & \Var^{0}_r & \rTo^{\cE_0} &
\Var^{1}_{r} & \rTo^{\cE_{1}} & \dots  & \rTo^{\cE_{n}} &
\Var^{n+1}_{r}  & \rTo^{\cE_{n+1}} & \cE_{n+1}(\Var^{n+1}_{r})  & \rTo^{\cE_{n+2}} & 
0 \,,
\end{diagram}
\eeq
where the sheaves $\Var^{p}_{r}$ with $0\leq p\leq n+2$ can be 
conveniently represented as shown in \cite{Vit98}. 
By means of this representation,
in \cite{FPV98a} we gave a suitable formulation of Noether's Theorem 
which will be useful in the sequel.

\bRm
We recall that a section $\lam\in\Var{n}_r$ is just a Lagrangian of order $(r+1)$ of 
the standard literature. 
Furthermore, 
$\cE_{n}(\lam) \in \Var^{n+1}_{r}$ coincides with the standard higher 
order Euler--Lagrange morphism associated with $\lam$.
\END
\eRm

\section{Gauge--natural Lagrangian theories}

We recall that in {\em gauge--natural} physical theories the fields are 
assumed to be sections of gauge--natural bundles, so that
right--invariant automorphisms of the structure bundle $\bP$ define 
uniquely the transformation laws of the fields themselves (see {\em e.g.} 
\cite{Ec81,KMS93}).
In the following, we shall develop a geometrical setting which enables us to 
define and investigate the physically fundamental concept of 
conserved quantity in gauge--natural theories.

\subsection{Gauge--natural prolongations}

First we shall recall some basic definitions and properties concerning
gauge--natural prolongations of (structure) principal bundles.

Let $\bP\to\bX$ be a principal bundle with structure group $\bG$.
Let $r\leq k$ and $\bW^{(r,k)}\bP$ $\byd$ $J_{r}\bP\ucar{\bX}L_{k}(\bX)$, 
where $L_{k}(\bX)$ is the bundle of $k$--frames 
in $\bX$ \cite{Ec81,Ja90,KMS93}, $\bW^{(r,k)}\bG \byd J_{r}\bG\odot GL_{k}(n)$
the semidirect product with respect to the action of $GL_{k}(n)$ 
on $J_{r}\bG$ given by the 
jet composition and $GL_{k}(n)$ is the group of $k$--frames 
in $\R^{n}$.

Elements of $\bW^{(r,k)}\bP$ are given by $(j^{\bx}_{r}\gam,j^{0}_{k}t)$, 
with $\gam: \bX\to\bP$ a local section, $t:\R^{n}\to\bX$ locally invertible 
at zero, 
with $t(0) = \bx$, $\bx\in \bX$. Elements of $\bW^{(r,k)}\bG$ are
$(j^{0}_{r}g,j^{0}_{k}\alp)$, where
$g:\R^{n}\to\bG$, $\alp: \R^{n}\to\R^{n}$ locally 
invertible at zero, with $\alp(0) = 0$.  

\bRm
$\bW^{(r,k)}\bP$ is a principal bundle over $\bX$ with structure group
$\bW^{(r,k)}\bG$.
The right action of $\bW^{(r,k)}\bG$ on the fibers of $\bW^{(r,k)}\bP$
is defined by the composition of jets (see, {\em e.g.}, 
\cite{Ja90,KMS93}).
\END\eRm

\bDf
The principal bundle $\bW^{(r,k)}\bP$ 
is said to be the {\em gauge--natural prolongation of order $(r,k)$ 
of $\bP$}. $\bW^{(r,k)}\bG$ is said to be the 
{\em gauge--natural prolongation of 
order $(r,k)$ of $\bG$}.
\END\eDf

\bRm
Let $(\Phi,\phi)$ be a principal automorphism of $\bP$ \cite{KMS93}.
It can be prolonged in a
natural  way to a principal automorphism of
$\bW^{(r,k)}\bP$, defined by:
\beq
\bW^{(r,k)}(\Phi, \phi):(j_{r}^{\bx}\gam, j_{k}^{0}t) \mto 
(j^{\phi(\bx)}_{r}(\Phi \circ \gam \circ \phi^{-1}),j^{0}_{k}(\phi \circ 
t))\,.
\eeq

The induced automorphism $\bW^{(r,k)}(\Phi,\phi)$ is an equivariant 
automorphism of $\bW^{(r,k)}\bP$ with 
respect to the action of the structure group $\bW^{(r,k)}\bG$. 
We shall simply denote it by the same symbol $\Phi$,  
if there is no danger of confusion.
\END\eRm

\bDf
We define the {\em vector}
bundle over $\bX$ of right--invariant infinitesimal automorphisms of $\bP$
by setting $\cA = T\bP/\bG$. 

We also define the {\em vector} bundle  over $\bX$ of right invariant 
infinitesimal automorphisms of $\bW^{(r,k)}\bP$ by setting 
$\cA^{(r,k)} \byd T\bW^{(r,k)}\bP/\bW^{(r,k)}\bG$ ($r\leq k$).
\END\eDf

\bRm
We have the following projections
\beq
\diagramstyle[size=2.3em] 
\begin{diagram}
& \cA^{(r,k)} &\rTo & \cA^{(r',k')}\,,\qquad r\leq k\,,\quad r'\leq k'\,,
\end{diagram}
\eeq
with $r\geq r'$, $s\geq s'$.
\END
\eRm

\bEx
Let $(\bP,\bX,\pi;\bG)$ be a principal bundle and 
$(U_{\alp},\gam^{(\alp)})$ a trivialization. The infinitesimal 
generator of the flow $(\Phi_{t},\phi_{t})$ is then given by
\bEq\label{Xi}
\Xi = \xi^{\mu}(\bx)\der_{\mu} + \xi^{A}(\bx)\rho_{A}\,,
\eEq
where $\rho_{A}$ is a pointwise basis of vertical 
right--invariant vector fields on 
$\bP$ and $\bx \in \bX$.
Here
$\xi^{\mu}(\bx) =
\frac{d}{dt}  [(\phi)_{t}(\bx)]_{t=0}$, $\xi^{A}(\bx) = 
(T^{-1})^{A}_{a}\frac{d}{dt} 
[(\Phi)^{a}_{t}(\bx)]_{t=0}$, $(\Phi_{t}; \phi_{t}) \in 
\text{Aut}(\bP)$ being a $1$--parameter flow of automorphisms of 
$\bP$ and $T_{A}=T^{a}_{A}\der_{a}^{(e)} \in T_{e}\bG$ being a basis 
of the Lie algebra $\mathfrak{g} \equiv T_{e}\bG$.
\END
\eEx

\bEx\label{esempio}
Let $\Xi$ be a right invariant vector field on $\bP$ with structure 
group $\bG=GL(m)$.
The coordinate expression of the gauge--natural prolonged 
right--invariant vector field $\bar{\Xi}$ in $\cA^{(1,1)}$ are given 
by (see \cite{Fa99})
\bEq\label{lift di xi}
\bar{\Xi}=\xi^{\mu}\der_{\mu}+\xi^{a}_{c}\rho^{c}_{a}+
\der_{\nu}\xi^{a}_{c}\rho^{c\nu}_{a}+
\der_{\nu}\xi^{\mu}\rho^{\nu}_{\mu}\,,
\eEq
where $\rho^{c}_{a} = g^{c}_{b}\der^{a}_{b}+ 
g^{c}_{b\nu}\der^{b\nu}_{a}$, $\rho^{c\nu}_{a}=g^{c}_{b}\der^{b\nu}_{a}$, 
$\rho^{\nu}_{\mu}=\eps^{\nu}_{\alp}\der^{\alp}_{\mu}-
g^{a}_{b\mu}\der^{b\nu}_{a}$ are a basis of vertical 
right--invariant vector fields on $\bW^{(1,1)}\bP$ while 
$(x^{\mu},g^{a}_{b},g^{a}_{b\mu})$ and $(x^{\mu}, \eps^{\mu}_{\alp})$ 
are local coordinates on $J_{1}\bP$ and $L(\bX)$, respectively.
\END
\eEx

\subsection{Gauge--natural bundles}

We recall that gauge--natural bundles are always 
of a finite order $(r,k)$ \cite{Ec81}. They are 
(isomorphic to) bundles associated to principal prolongations 
of principal bundles, so 
that the study of gauge--natural bundles can be reduced to the 
representation theory of the Lie groups $\bW^{(r,k)}\bG$ 
(see \cite{Ec81} and references quoted in \cite{KMS93}).

Let $\bF$ be any manifold and $\zet:\bW^{(r,k)}\bG\ucar{}\bF\to\bF$ be 
a left action of $\bW^{(r,k)}\bG$ on $\bF$. There is a naturally defined 
rigth action of $\bW^{(r,k)}\bG$ on $\bW^{(r,k)}\bP \times \bF$ so that
 we can associate in a standard way
to $\bW^{(r,k)}\bP$ the bundle, on the given basis $\bX$,
$\bY_{\zet} \byd \bW^{(r,k)}\bP\times_{\zet}\bF$.

\bDf
We say $(\bY_{\zet},\bX,\pi_{\zet};\bF,\bG)$ to be the 
{\em gauge-natural bundle} of order 
$(r,k)$ associated to the principal bundle $\bW^{(r,k)}\bP$ 
by means of the left action $\zet$ of the group 
$\bW^{(r,k)}\bG$ on the manifold $\bF$. 
\END\eDf

\bRm
A principal automorphism $\Phi$ of $\bW^{(r,k)}\bP$ induces an 
automorphism of the gauge--natural bundle by:
\bEq
\Phi_{\zet}:\bY_{\zet}\to\bY_{\zet}: [(j^{x}_{r}\gam,j^{0}_{k}t), 
\hat{f}]_{\zet}\mto [\Phi(j^{x}_{r}\gam,j^{0}_{k}t), 
\hat{f}]_{\zet}\,, 
\eEq
where $\hat{f}\in \bF$ and $[\cdot, \cdot]_{\zet}$ is the equivalence class
induced by the action $\zet$.
\END\eRm

\bEx
The associated bundle $\bW^{(1,1)}\bP\times_{\mu}
(\R^{n}\oplus \mathfrak{g})$, 
where $\mu$ is the natural action of $J_{1}\bG \odot GL(n)$ over 
$(\R^{n}\oplus \mathfrak{g})$, is a gauge--natural bundle of order $(1,1)$.
A given automorphism of the principal bundle $\bP$ acts on 
$\bW^{(1,1)}\bP\times_{\mu}\R^{n}\oplus 
\mathfrak{g}$ by 
means of the canonical action of $\text{Aut}(\bP)$.
Global section of $\bW^{(1,1)}\bP\times_{\mu}
(\R^{n}\oplus \mathfrak{g})$ are in a one--to--one 
correspondence with right--invariant infinitesimal automorphisms of $\bP$.
In fact it is easy to prove that $\bW^{(1,1)}\bP\times_{\mu}
(\R^{n}\oplus \mathfrak{g})$ is a {\em vector} bundle (see \cite{Fa99,Ja90}) and 
that the isomorphism 
$\bW^{(1,1)}\bP\times_{\mu}(\R^{n}\oplus \mathfrak{g})
\simeq \cA$ holds true.
\END\eEx

\subsection{Gauge--natural lift}\label{Lift}

Denote by $\cT_{\bX}$ and $\cA^{(r,k)}$ the sheaf of
vector fields on $\bX$ and the sheaf of right invariant vector fields 
on $\bW^{(r,k)}\bP$, respectively.

From now on we assume a functorial mapping is defined 
which lifts any right--invariant local automorphism $(\Phi,\phi)$ of the 
principal bundle $W^{(r,k)}\bP$ into a unique local automorphism 
$(\Phi_{\zet},\phi)$ of the associated bundle $\bY_{\zet}$. 
This lifting depends linearly on derivatives of $\xi^{A}$ and 
$\xi^{\mu}$ up to order $r$ and $k$, respectively.
Its infinitesimal version associates to any $\bar{\Xi} \in \cA^{(r,k)}$,
projectable over $\xi \in \cT_{\bX}$, a unique {\em projectable} vector field 
$\hat{\Xi} \byd \cN(\bar{\Xi})$ on $\bY_{\zet}$ in the 
following way:
\bEq
\cN : \bY_{\zet} \ucar{\bX} \cA^{(r,k)} \to T\bY_{\zet} \,:
(\by,\bar{\Xi}) \mto \hat{\Xi} (\by) \,,
\eEq
where, for any $\by \in \bY_{\zet}$, one sets: $\hat{\Xi}(\by)=
\frac{d}{dt} [(\Phi_{\zet \,t})(\by)]_{t=0}$,
and $\Phi_{\zet \,t}$ denotes the (local) flow corresponding to the 
gauge--natural lift of $\Phi_{t}$.

This mapping fulfils the following properties:
\begin{enumerate}
\item $\cN$ is linear over $id_{\bY_{\zet}}$;
\item we have $T\pi_{\zet}\circ\cN = id_{T\bX}\circ 
\bar{\pi}^{(r,k)}$, 
where $\bar{\pi}^{(r,k)}$ is the natural projection
$\bY_{\zet}\ucar{\bX} 
\cA^{(r,k)} \to T\bX$;
\item for any pair $(\bar{\Lam},\bar{\Xi})$ of vector fields in 
$\cA^{(r,k)}$, we have
\beq
\cN([\bar{\Lam},\bar{\Xi}]) = [\cN(\bar{\Lam}), \cN(\bar{\Xi})]\,.
\eeq
\item we have the coordinate expression of $\cN$
\bEq\label{espressione}
\cN = d^\mu \ten \der_\mu + d^{A}_{\bnu}
\ten (\cZ^{i\bnu}_{A} \der_{i}) + d^{\nu}_{\blam}
\ten (\cZ^{i\blam}_{\nu} \der_{i}) \,,
\eEq
with $0<|\bnu|<k$, $1<|\blam|<r$ and 
$\cZ^{i\bnu}_{A}$, $\cZ^{i\blam}_{\nu}$ $\in C^{\infty}(\bY_{\zet})$ 
are suitable functions which depend on the bundle.
\end{enumerate}

\bDf
The map $\cN$ is called the {\em gauge--natural lifting 
functor}. 
The vector field $\hat{\Xi}\equiv \cN(\bar{\Xi})$ is 
called the {\em 
gauge--natural lift} of $(\bar{\Xi},\xi)$ to the bundle $\bY_{\zet}$.\END
\eDf


\subsection{Principal connections}

Let $(\bP,\bX,\pi;\bG)$ be a principal fiber bundle.
Vertical vector fields on $\bP$ are intrinsic objects, while the 
notion of horizontal vector fields on $\bP$ is given by means of a 
principal connection on $\bP$ (see\cite{KMS93}).

It is well known that a 
principal connection over $\bP$ can be represented  by means of a
$\mathfrak{g}$--valued $1$--form $\ome$ over $\bP$,
$\mathfrak{g}$ being the Lie 
algebra of the Lie group $\bG$. 
This form has, locally, the 
coordinate expression:
\beq
\ome = d^{\mu} \ten (\der_{\mu}+\ome^{A}_{\mu}(\bx)\rho_{A})\,.
\eeq

A principal connection on $\bP$ induces a splitting of the exact 
sequence of vector bundles over $\bX$:
\beq
\diagramstyle[size=2.3em]
\begin{diagram}
0 &\rTo & V\bP/\bG &\rTo & \cA & \rTo & T\bX & \rTo & 0 \,,
\end{diagram}
\eeq
so that we have the (non--canonical) isomorphism 
$\cA \simeq V\bP/\bG \oplus T\bX$.

Let $\ome$ be a principal connection on $\bP$. 
An infinitesimal automorphism of $\bP$ is then given 
in the splitted form 
\bEq\label{sp}
\Xi = \Xi_{h} + \Xi_{v}= \xi^{\mu}(\der_{\mu} + 
\ome^{A}_{\mu}\rho_{A}) + (\xi^{A} - 
\ome^{A}_{\mu}\xi^{\mu})\rho_{A} \label{splitt}\,.
\eEq
We shall respectively indicate by $\Xi_{v}$ and $\Xi_{h}$ 
the vertical and horizontal components of $\Xi$ with 
respect to the principal connection $\ome$; furthermore we shall write 
$\xi^{A}_{v} = \xi^{A} - \ome^{A}_{\mu}\xi^{\mu}$.

\medskip

By a straightforward but tedious coordinate calculus it is easy to 
verify the following (see also \cite{Kol75}, 
\cite{Kol79} Proposition $4$, and references quoted therein).

\bLm
Let $\ome$ be a principal connection on $\bP$ and $\Gam$ 
a linear symmetric connection on $L(\bX)$.
Let $\bar{\ome}\equiv \bW^{(r,k)}(\ome,\Gam)$ be the connection induced 
on $\bW^{(r,k)}\bP$ by the 
gauge--natural prolongation functor. 
Then 
$\bar{\ome}$ is a principal connection on $\bW^{(r,k)}\bP$.
\eLm

Then we have the following:
\bPr\label{main result}
Let $\ome$ be a principal connection on $\bP$ and $\Gam$ 
a linear symmetric connection on $L(\bX)$ such that $\bar{\ome}$ is the prolongation of $\ome$ 
with respect to $\Gam$ in the sense of \cite{Kol75}.
A global, non--canonical, isomorphism holds:
\bEq\label{ISO}
\cA^{(r,k)}\simeq_{(\ome,\Gam)} J_{r}(V\bP/\bG)\ucar{\bX} J_{k}T\bX\,.
\eEq 
\ePr

\bPf
The gauge--natural lift of the splitting of $\Xi$ by means 
of $\ome$ corresponds to the splitting of $\bar{\Xi}$ by 
means of the induced principal connection $\bar{\ome}$ on $\bW^{(r,k)}\bP$ 
depending on $(\ome,\Gam)$. This specifies the isomorphism.
\QED
\ePf

\subsection{Lie derivative}

Let $\gam$ be a (local) section of $\bY_{\zet}$, $\bar{\Xi}$ 
$\in \cA^{(r,k)}$ and $\hat\Xi$ its gauge--natural lift. 
Following \cite{KMS93} we
define a (local) section $\pounds_{\bar{\Xi}} \gam : \bX \to V\bY_{\zet}$, 
by setting:
$\pounds_{\bar{\Xi}} \gam = T\gam \circ \xi - \hat{\Xi} \circ \gam$. 

\bDf
The (local) section $\pounds_{\bar{\Xi}} \gam$ is called the {\em 
generalized Lie derivative} of $\gam$ along the vector field 
$\hat{\Xi}$.
\END\eDf

\bRm
This section is a vertical prolongation of $\gam$, \ie it satisfies
the property: $\nu_{\bY_{\zet}} \circ \pounds_{\Xi} \gam = \gam$, 
where $\nu_{\bY_{\zet}}$ is the projection 
$\nu_{\bY_{\zet}}: V\bY_{\zet} \to \bY_{\zet}$.
Its coordinate expression is given by
$(\pounds_{\bar{\Xi}}\gam)^{i} = \xi^{\sig} \der_{\sig} \gam ^{i} - 
\hat\Xi^{i}$.

Furthermore, we can consider $\pounds$ as a bundle morphism
\bEq
\pounds: J_{1}\bY_{\zet} \ucar{\bX}
\cA^{(r,k)} \to V\bY_{\zet}\,.
\eEq
\END\eRm

\bRm\label{lie}
The Lie derivative operator satisfies the following 
properties:
\begin{enumerate}\label{lie properties}
\item for any vector field $\bar{\Xi} \in \cA^{(r,k)}$, the 
mapping $\gam \mto \pounds_{\bar{\Xi}}\gam$ 
is a first--order quasilinear differential operator;
\item for any local section $\gam$ of $\bY_{\zet}$, the mapping 
$\bar{\Xi} \mto \pounds_{\bar{\Xi}}\gam$ 
is a linear differential operator depending on the derivatives of 
$\xi^{A}$ up to order $r$ and on the derivatives of $\xi^{\mu}$ up to order 
$k$, respectively;
\item by using the canonical 
isomorphism $VJ_{r}\bY_{\zet}\simeq J_{r}V\bY_{\zet}$, we have
$\pounds_{\bar{\Xi}}[j_{r}\gam] = j_{r} [\pounds_{\bar{\Xi}} \gam]$,
for any (local) section $\gam$ of $\bY_{\zet}$ and for any (local) 
vector field $\bar{\Xi}\in \cA^{(r,k)}$.
\end{enumerate}
\END\eRm

\section{Gauge--natural Lagrangians and their symmetries}

We consider now a projectable vector field $(\hat{\Xi},\xi)$ on 
$\bY_{\zet}$ and take into account the Lie derivative with respect to 
its prolongation $j_{s}\hat{\Xi}$. Such a prolonged vector field 
preserves the fiberings; hence it preserves the natural 
splitting \eqref{jet connection}. 

It is known \cite{FPV98a} that the standard 
Lie derivative operator with respect to the r-th prolongation 
$j_{s}\hat{\Xi}$ of a projectable vector field $(\hat{\Xi},\xi)$
on $\bY_{\zet}$ passes to the quotient spaces 
$\For^{p}_{s}/\Thd^{p}_{s}$ 
so that it can be represented by an operator 
(the {\em variational Lie derivative\/}) 
on the sheaves of the short variational sequence associated with 
$\bY_{\zet}$. In the case $p = n$ we have in particular
\bEq \label{noether}
\cL_{j_{s}\hat{\Xi}}: \Var^{n}_{s} \to \Var^{n}_{2s+1}: \lam \mto 
\hat{\Xi}_{V}\rfloor \cE(\lam)+
d_{H}(j_{s}\bar{\Xi}_{V}\rfloor p_{d_{V}\lam}+\xi\rfloor \lam) \,,
\eEq
where $p_{d_{V}\lam}$ is a momentum associated to $\lam$ (see 
\cite{Fe84,FPV98a,Vit98}) and the subscript $V$ here means the vertical 
component with respect to the natural splitting \eqref{jet connection}.

Variational Lie derivatives allow us to compute 
\lq variationally relevant\rq \ infinitesimal symmetries 
of Lagrangians in the variational sequence. 

\bDf\label{gn}
Let $(\hat{\Xi},\xi)$ be a projectable vector field on $\bY_{\zet}$. 
Let $\lam \in \Var^{n}_{s}$ 
be a generalized Lagrangian. We say $\hat{\Xi}$ to be a {\em symmetry\/} 
of $\lam$ if $\cL_{j_{s+1}\hat{\Xi}}\,\lam = 0$.

We say $\lam$ to be a 
{\em gauge--natural Lagrangian} if the lift 
$(\hat{\Xi},\xi)$ of any vector 
field $\bar{\Xi} \in \cA^{(r,k)}$ is a  symmetry for 
$\lam$, {\em i.e.} if $\cL_{j_{s+1}\bar{\Xi}}\,\lam = 0$. 
In this case the projectable vector field 
$\hat{\Xi}\equiv \cN(\bar{\Xi})$ is 
called a {\em gauge--natural symmetry} of $\lam$. \END 
\eDf

\bRm
We can regard $\pounds_{\bar{\Xi}}: J_{1}\bY_{\zet} \to V\bY_{\zet}$ 
as a morphism over the
basis $\bX$. In this case it is meaningful to consider the (standard)
prolongation of $\pounds_{\bar{\Xi}}$, denoted by 
$j_{s}\pounds_{\bar{\Xi}}: 
J_{s+1}\bY_{\zet} \to VJ_{s}\bY_{\zet}$, where we made use 
of the isomorphism $(iii)$ recalled in Remark \ref{lie}.
\END\eRm

Symmetries of a Lagrangian $\lam$ are calculated by means of 
Noether's Theorem, which takes a particularly interesting 
form in the case of gauge--natural Lagrangians.  

\bTh (Noether's Theorem for gauge--natural Lagrangians)
\label{symmetry of L}
Let $\lam \in \Var^{n}_{s}$ be a gauge-natural Lagrangian and 
$(\hat{\Xi},\xi)$ 
a gauge--natural symmetry of $\lam$. Then by \eqref{noether} we have
\bEq\label{first decomposition}
0 = - \pounds_{\bar{\Xi}} \rfloor \cE(\lam) 
+d_{H}(-j_{s}\pounds_{\bar{\Xi}} 
\rfloor p_{d_{V}\lam}+ \xi \rfloor \lam) \,.
\eEq
Suppose that the section $\sig:\bX \to \bY_{\zet}$ fulfills the 
condition
\bEq
(j_{2s+1}\sig)^{*}(- \pounds_{\bar{\Xi}} \rfloor \cE(\lam)) = 0 \,.
\eEq
Then, the $(n-1)$--form 
\bEq
\eps = - j_{s}\pounds_{\bar{\Xi}} \rfloor p_{d_{V}\lam}+ \xi 
\rfloor \lam \label{current}\,,
\eEq
fulfills the equation $d ((j_{2s}\sig)^{*}(\eps)) = 0$.
\eTh

\bRm
If $\sig$ is a critical section for $\cE(\lam)$, \ie
$(j_{2s+1}\sig)^{*}\cE(\lam) = 0$, the above equation 
admits a physical interpretation as a {\em weak conservation law} 
for the density associated with $\eps$.
\END\eRm

\bDf
Let $\lam \in \Var^{n}_{s}$ be a gauge--natural Lagrangian and 
$\bar{\Xi} \in \cA^{(r,k)}$. Then the sheaf morphism $\eps$
is said to be a {\em gauge--natural conserved current\/}.\END 
\eDf

\bRm\label{arbitrary1}
In general, this conserved current is not uniquely defined. In fact, 
it depends on the choice of $p_{d_{V}\lam}$, which is not unique (see 
\cite{Vit98} and references quoted therein).
Moreover, we could add to the conserved current any form
$\bet \in \Var^{n-1}_{2s}$ which is variationally closed, \ie such 
that $\cE_{n-1}(\bet) = 0$ holds. The form $\bet$ is 
locally of the type $\bet = d_{H}\gam$, where $\gam \in 
\Var^{n-2}_{2s+1}$.
\END\eRm


Proposition \ref{main result} enables us to state a technical lemma 
which will be useful in the sequel.

\bLm\label{Fund}
Let $\alp: J_{s}(\bY_{\zet} \ucar{\bX} \cA^{(r,k)})$ $\to$ $\owed{p}T^{*}\bX$ 
be a linear morphism with respect to the 
fibering $J_{s}\bY_{\zet} \ucar{\bX}J_{s}\cA^{(r,k)}$ $\to$ 
$J_{s}\bY_{\zet}$ and let $D_{H}$ be the horizontal differential 
on $\bY_{\zet} \ucar{\bX} \cA^{(r,k)}$. We can uniquely write $\alp$ as
\beq
\olin{\alp}: J_{s}\bY_{\zet} \to (\cC^{*}_{s+r}[(V\bP/\bG)]\ucar{\bX} 
\cC^{*}_{s+k}[T\bX])\wed (\owed{p}T^{*}\bX)\,.
\eeq
Then
\bEq\label{tool}
\olin{D_{H}\alp} = D_{H}\olin{\alp}\,.
\eEq
\eLm

\bPf
This is a naturality property of $D_{H}$ which follows from the
linearity of $D_{H}\alp$ with respect to the fibering 
$J_{s+1}\bY_{\zet}\ucar{\bX} J_{s+1}\cA^{(r,k)} 
\to J_{s+1}\bY_{\zet}$, 
the non--canonical isomorphism \eqref{ISO}
and the isomorphisms 
$J_{s}T\bX\ucar{\bX}(J_{s}T\bX)^* \simeq V^{*}J_{s}T\bX$, 
$J_{s}(V\bP/\bG)\ucar{\bX}(J_{s}(V\bP/\bG))^* \simeq V^{*}J_{s}(V\bP/\bG)$,
$\Con^{*}_{s+1}[T\bX]$ $\wed$ $\owed{p+1}T^{*}\bX$ $\simeq$ $V^{*}J_{s+1}T\bX$ 
$\ten$ $\owed{p+1}T^{*}\bX$, 
$\Con^{*}_{s+1}[(V\bP/\bG)]$ $\wed$ $\owed{p+1}T^{*}\bX$ $\simeq$ 
$V^{*}J_{s+1}(V\bP/\bG)$ $\ten$ $\owed{p+1}T^{*}\bX$.\QED
\ePf

We stress that the above result depends on a non--canonical isomorphism 
and specifically on the choice of $(\ome,\Gam)$.

\bRm\label{horizontal diff}
Let $\eps : J_{2s}\bY_{\zet} \ucar{\bX} \cA^{(r,k)} 
\to \owed{n-1} T^{*}\bX$ be a conserved current. 
As an immediate consequence of the above lemma we have that, making 
use of \eqref{tool}, we can regard $\eps$ as the 
equivalent morphism $\olin{\eps}: J_{2s}\bY_{\zet} \to (\cC_{r}^{*}[(V\bP/\bG)]
\ucar{\bX}  \cC^{*}_{k}[T\bX]) 
\wed (\owed{n-1} T^{*}\bX)$.
\END\eRm


\section{Superpotentials}

As is well known in gauge--natural lagrangian theories 
\cite{Fa99}, 
performing covariant integrations by 
parts enables us to decompose the current $\eps$ into the sum of 
the so--called {\em reduced current} and the formal 
divergence of a skew--symmetric tensor density called {\em 
superpotential} (which is defined modulo a divergence). 
Conservation laws which occur in 
gauge--natural theories are {\em strong laws}, \ie they hold along 
any (not necessarily critical) section of the bundle. Along critical sections 
the reduced current vanishes identically so that the current $\eps$ 
is not only closed, but it is also exact along solutions of the 
Euler--Lagrange equations.

\bRm
Let $\lam$ be a gauge--natural Lagrangian. By the linearity of
$\pounds$ with respect to the vector bundle structure 
$J_{r}(V\bP/\bG)\ucar{\bX}J_{k}T\bX 
\to \bX$ we have 
\beq
\mu \equiv \mu (\lam) = \pounds \rfloor \cE (\lam): J_{2s}\bY_{\zet}
\to (\cC_{r}^{*}[(V\bP/\bG)]\ucar{\bX}\cC^{*}_{k}[T\bX])\wed (\owed{n} 
T^{*}\bX) \,.\END
\eeq
\eRm

In the following we shall give the main result which enables us to 
describe gauge--natural superpotentials in the short variational 
sequence. We shall apply to the `total' space $\bY_{\zet} \ucar{\bX} 
(V\bP/\bG)\ucar{\bX} T\bX$ 
a standard result concerning the integration by parts procedure 
involved in variational formulae (see {\em e.g.} 
\cite{FPV98a,FPV98b,Vit98}).

The following Lemma is an application of an
abstract result due to Kol\'a\v r and Hor\'ak \cite{HoKo83,Kol84} 
concerning a global decomposition formula for suitable morphisms.

\bLm\label{kol}
Let $\mu: J_{2s}\bY_{\zet} \to (\cC_{r}^{*}[(V\bP/\bG)] 
\ucar{\bX}  \cC^{*}_{k}[T\bX])\wed
(\owed{p}T^{*}\bX)$,
with $0\leq p \leq n$ and let $D_{H}\mu = 0$.
We regard $\mu$ as the extended morphism 
$\hat{\mu}: J_{2s}(\bY_{\zet} \ucar{\bX} \cA^{(r,k)})
\to (\cC^{*}_{r}[\bY_{\zet} \ucar{\bX} (V\bP/\bG)]
\ucar{\bX}  \cC^{*}_{k}[T\bX])\wed
(\owed{p}T^{*}\bX)$. 
Then we have globally
\beq
\hat{\mu} = E_{\hat{\mu}} + F_{\hat{\mu}}\,,
\eeq 
where
\beq
E_{\hat{\mu}}: J_{2s}(\bY_{\zet} \ucar{\bX} \cA^{(r,k)}) \to 
(\cC^{*}_{0}[\bY_{\zet} \ucar{\bX} (V\bP/\bG)]\ucar{\bX}  \cC^{*}_{0}[T\bX])\wed
(\owed{p}T^{*}\bX) \,,
\eeq
locally, $F_{\hat{\mu}} = D_{H}M_{\hat{\mu}}$, with
\beq
M_{\hat{\mu}}: J_{2s-1}(\bY_{\zet} \ucar{\bX} \cA^{(r,k)}) \to 
(\cC^{*}_{r-1}[\bY_{\zet} \ucar{\bX} (V\bP/\bG)]
\ucar{\bX}  \cC^{*}_{k-1}[T\bX])\wed 
(\owed{p-1}T^{*}\bX) \,.
\eeq
\eLm

\bPf
$E_{\hat{\mu}}$ and $D_{H}M_{\hat{\mu}}$ can be evaluated by
means of backwards procedures (see {\em e.g.} \cite{Fe84,Kol83}).\QED
\ePf

\bRm\label{arbitrary2}
In general there is no uniquely determined $M_{\hat{\mu}}$. 
In fact it can be proved that a linear symmetric connection on 
$\bX$ yields a distinguished choice of $M_{\hat\mu}$ in analogy to 
\cite{Kol83}, Prop. $1$; see also \cite{Fe84,FFR84}.
\END
\eRm

\bTh
Let $\hat{\mu}$ be of the type $\mu: J_{2s}\bY_{\zet} \to 
(\cC_{r}^{*}[(V\bP/\bG)] \ucar{\bX}  \cC^{*}_{k}[T\bX])\wed
(\owed{n}T^{*}\bX)$, then the following decomposition formula holds
\bEq
\mu = \tilde{\mu} + D_{H}\tilde{\eps} \,,
\eEq
where
\beq
\tilde{\mu} \byd E_{\mu}: J_{2s}\bY_{\zet} \to (\cC_{r}^{*}[(V\bP/\bG)] 
\ucar{\bX}  \cC^{*}_{k}[T\bX])\wed \owed{n}T^{*}\bX \,,
\eeq
and
\beq
\tilde{\eps} \byd M_{\mu}: J_{2s-1}\bY_{\zet} \to 
(\cC^{*}_{r-1}[(V\bP/\bG)] \ucar{\bX}  \cC^{*}_{k-1}[T\bX])
\wed(\owed{n-1}T^{*}\bX) \,.
\eeq
\eTh
 
\bPf
We take into account that $D_{H}\mu$ is obviously vanishing, then the 
result is a straightforward consequence of Lemma \ref{kol} with $p = 
n$.\QED
\ePf

\bRm\label{bianchi}
For any $(\Xi,\xi)$, the morphism $\tilde{\mu} \equiv \cE(\mu((\Xi,\xi)))$
is identically vanishing. So, we have $\mu = D_{H}\tilde{\eps}$.
We stress that these are just the {\em generalized Bianchi 
identities}.\END
\eRm

\bDf 
The form $\tilde{\eps}$ is said to be a {\em reduced current}.\END  
\eDf

\bRm
If the coordinate expression of $\mu$ is given by
\beq
\mu= (\mu^{\balp}_{i}\vartht^{i}_{\balp}+
\mu^{\bA}_{i}\vartht^{i}_{\bA}) \wed \ome \,,
\eeq
with $\vartht^{i}_{\bA}$ and $\vartht^{i}_{\balp}$ contact forms 
on $J_{r}(V\bP/\bG)$ and $J_{k}T\bX$ respectively, 
then the coordinate expression of $\tilde{\eps}$ is given by
\beq
\tilde{\eps} = (\tilde{\eps}_{i}^{\balp +\sig}\vartht^{i}_{\balp} +
\tilde{\eps}_{i}^{\bA +\sig}\vartht^{i}_{\bA})\wed \ome_{\sig}\,,
\eeq
where $\tilde{\eps}_{i}^{\balp +\sig}$ and $\tilde{\eps}_{i}^{\bA +\sig}$ 
are (not uniquely) determined in terms of $\mu^{\balp}_{i}$ and 
$\mu^{\bA}_{i}$ (see {\em e.g.} \cite{FeFr91,Kol83}).\END
\eRm

\bCr
Let $\lam \in \Var^{n}_{s}$ be a gauge--natural Lagrangian and 
$(\hat{\Xi},\xi)$ 
a gauge--natural symmetry of $\lam$ according to Definition \ref{gn}. 
Then, being $\mu = D_{H}\eps$,
the following holds by virtue of Remark \ref{bianchi}:
\bEq\label{strong conservation}
D_{H}(\eps -\tilde{\eps}) = 0\,.
\eEq
\eCr

Eq. \eqref{strong conservation} is referred as a gauge--natural 
`strong conservation law' for the density $\eps -\tilde{\eps}$.

We can now state the following main result about the existence (and 
globality) of gauge--natural superpotentials in the framework 
of variational sequences.

\bTh\label{global}
Let $\lam \in \Var^{n}_{s}$ be a gauge--natural Lagrangian and 
$(\hat{\Xi},\xi)$ a gauge--natural symmetry of $\lam$. Then there exists 
a (global) sheaf morphism 
$\eta \in \left(\Var^{n-2}_{2s-1}\right)_{\bY_{\zet} \ucar{\bX} 
\cA^{(r,k)}}$
such that
\beq
D_{H}\eta = \eps -\tilde{\eps}\,.
\eeq
\eTh

\bPf
\begin{enumerate}
\item (local existence)
By applying Lemma \ref{Fund}, we can consider 
\beq
\eps -\tilde{\eps}: J_{2s-1}\bY_{\zet} \ucar{\bX} 
\cA^{(r,k)} \to \owed{n-1}T^{*}\bX\,, 
\eeq
then we take eq. \eqref{strong conservation} into account and we 
integrate over the variational sequence associated with 
$\bY_{\zet} \ucar{\bX} \cA^{(r,k)}$. 
\item (global existence)
Eq. \eqref{strong conservation} assures us that the hypotheses of 
Lemma \ref{kol} are satisfied, so we have
\beq
\eps -\tilde{\eps}=\olin{\eps -\tilde{\eps}} + D_{H}\eta\,,
\eeq
where $\olin{\eps -\tilde{\eps}}$ is vanishing because of a uniqueness 
argument. Globality follows from Lemma \ref{kol} for $p=n-1$.
\end{enumerate}
\QED
\ePf

\bDf
We define the sheaf morphism $\eta$ to be a {\em gauge-natural 
superpotential\/} of $\lam$.
\END\eDf

\bRm 
As a consequence of Remarks \ref {arbitrary1} and \ref{arbitrary2},
superpotentials are not defined uniquely. In fact the 
choice of linear symmetric connections over $\bX$ generally yields 
distinguished superpotentials.\END
\eRm

\bRm
Theorem \ref{global} is based essentially on the 
vector bundle structure of $\cA^{(r,k)}$. 
It is noteworthy that the non--canonical splitting \eqref{ISO} 
induces a decomposition of 
variational objects like conserved currents and superpotentials. This 
decomposition is not natural (see \cite{Ec81}) and the choice of a 
{\em dynamical} connection, \ie a connection depending on physical 
fields together with their derivatives, turns out to be the physically 
most significant one.

We shall respectively denote by
\bEq
\eps_{h} = - j_{s}\pounds_{\bar{\Xi}_{h}} \rfloor p_{d_{V}\lam}+ \xi 
\rfloor \lam \label{currenth}\,,
\eEq
the ``horizontal'' (\ie ``natural'') part and by
\bEq
\eps_{v} = - j_{s}\pounds_{\bar{\Xi}_{v}} \rfloor p_{d_{V}\lam}+ \xi 
\rfloor \lam \label{currentv}\,,
\eEq
the ``vertical'' (\ie ``gauge'') part of the conserved current 
$\eps$ with respect to any non--canonical splitting.

We shall respectively denote by $\eta_{h}$ and by $\eta_{v}$  the 
``horizontal'' (\ie ``natural'') and the ``vertical'' (\ie ``gauge'') 
part of the superpotential with respect to any non--canonical splitting.
\END
\eRm
\subsection{An example of application}

In this Section we shall show how the formalism developed here enables 
us to obtain in a very straightforward way
well known results concerning conserved quantities in
the case of Einstein--Yang--Mills theories (see {\em e.g.} \cite{Fa99,GMV91} 
and references therein).

\medskip

Let $\lam\in\Var{n}_r$.
Then the following coordinate expressions hold:
\beq
d_{V}\lam = (d_{V}\lam)^{\balp}_{i}\vartht^{i}_{\balp}\wed \ome \,,
\,
E_{d_{V}\lam}
= \cE(\lam)_{i}\vartht^{i}\wed \ome \,,
\,
p_{d_{V}\lam}
= p(\lam)^{\balp\mu}_{i}\vartht^{i}_{\balp}\wed \ome_{\mu}\,.
\eeq

It is known (see {\em e.g.} \cite{Kol83}) that
\begin{align}
p(\lam)^{\bbet\mu}_{i} 
&= (d_{V}\lam)^{\balp}_{i} \qquad \bbet +\mu= \balp, |\balp|= r \label{k} \,,
\\
p(\lam)^{\bbet\mu}_{i}
&= (d_{V}\lam)^{\balp}_{i} - D_{\nu}p(\lam)^{\balp \nu}_{i}
\qquad \bbet+\mu=\balp, |\balp|= r-1 \label{kkk} \,,
\\
\cE(\lam)^{\balp}_{i}
&= (d_{V}\lam)^{\balp}_{i}- D_{\nu}p(\lam)^{\balp\nu}_{i}\qquad |\balp|=0
\label{kk}\,.
\end{align}
Furthermore,
$\cE(\lam)_{i}=\sum_{|\balp|\leq r}(-1)^{|\balp|}
D_{\balp}(d_{V}\lam)^{\balp}_{i}$.

\medskip

Let $(\bP,\bX,\pi;\bG)$ be a principal bundle, $g$ a metric on 
$\bX$, $k$ an $ad$--invariant metric on $\bG$. 
Let $\ome$ be a principal connection and $F$ its $\mathfrak{g}$--valued 
curvature $2$--form.

Let us now take the {\em gauge--natural} bundle 
$\bY = Lor(\bX)\ucar{\bX}\bC$, where $Lor(\bX)$ is the
bundle of Lorentzian metrics 
over space--time $\bX$ and $\bC$ is the affine 
bundle of principal connections $\ome$ over $\bP$,
whose associated vector bundle is the tensor product bundle 
$T^{*}\bX \ten V\bP/\bG$ \cite{GMV91}. Local coordinates on $\bY$ 
are given by $x^{\mu}, g^{\mu\nu},\ome^{A}_{\mu}$.
 
Let us consider the {\em gauge--natural} Lagrangian $\lam$ defined on 
the gauge--natural bundle $J_{2}Lor(\bX)\ucar{\bX}J_{1}\bC$:
\bEq
\lam=\lam_{H}(g^{\mu\nu},R_{\mu\nu})+\lam_{YM}(g^{\mu\nu},F^{A}_{\mu\nu})\,,
\eEq
where $\lam_{H}=-\frac{1}{2\kappa}\sqrt{g}g^{\alp\bet}R_{\alp\bet}$ is
the Einstein Lagrangian, $R_{\alp\bet}$ is the (formal) Ricci tensor of the
metric $g$ given by 
$R_{\alp\bet}\byd
R^{\mu}_{\alp\mu\bet}=D_{\mu}\gam^{\mu}_{\alp\bet}-D_{\bet}\gam^{\mu}_{\alp\mu}+
\gam^{\mu}_{\nu\mu}\gam^{\nu}_{\alp\bet}-
\gam^{\mu}_{\nu\bet}\gam^{\nu}_{\alp\mu}$, with
$\gam^{\mu}_{\nu\bet}=\frac{1}{2}g^{\mu\alp}(D_{\nu}g_{\bet\alp}-
D_{\alp}g_{\nu\bet}+D_{\bet}g_{\alp\nu})$ the (formal) Levi--Civita connection
of $g$, $\sqrt{g}=\sqrt{|det(g^{\mu\nu})|}$, $\kappa$ is a constant and
$\lam_{YM}(g_{\mu\nu},F^{A}_{\mu\nu})=
-\frac{1}{4}\sqrt{g}F^{\lam\gam}_{A}F_{\lam\gam}^{A}$ is the (gauge)
Yang--Mills Lagrangian. Here
$F^{\lam\gam}_{A}=k_{AB}g^{\lam\alp}g^{\gam\bet}F^{B}_{\alp\bet}$.

Notice that in this case $\bY_{\zet}= J_{2}Lor(\bX)\ucar{\bX}J_{1}\bC$
and $(r,k)=(3,2)$.

Let $\Xi$ be a generator of automorphisms of $\bP$.
From \eqref{currenth} and \eqref{currentv}, by means of \eqref{k}--\eqref{kk}, 
we get
\bEq
\eps^{\sig}(\lam,\Xi_{h})
=\eps^{\sig}(\lam_{H},\Xi_{h})+\eps^{\sig}(\lam_{YM},\Xi_{h})\,,
\eEq
where
\beq
\eps^{\sig}(\lam_{H},\Xi_{h})= 
\frac{1}{\kappa}\sqrt{g}(R^{\sig}_{\bet}-Rg^{\sig}_{\bet})\xi^{\bet}
+\nabla_{\mu}[\frac{\sqrt{g}}{2\kappa}
(\nabla^{\sig}\xi^{\mu}-\nabla^{\mu}\xi^{\sig})]\,.
\eeq
and 
\beq
\eps^{\sig}(\lam_{YM},\Xi_{h})=
(2p^{\mu\sig}_{A}\pounds_{\Xi_{h}}\ome^{A}_{\mu}-\lam_{YM}\xi^{\sig})=
-\sqrt{g}(F^{\mu\sig}_{A}F^{A}_{\mu\nu}-
\frac{1}{4}F^{\mu\rho}_{A}F^{A}_{\mu\rho}\del^{\sig}_{\nu})\xi^{\nu}\,.
\eeq
$R$ is the scalar curvature and 
$p^{\mu\nu}_{A}=-\frac{\sqrt{g}}{2}F^{\mu\nu}_{A}$. 
Here and in the sequel $\nabla_{\mu}$ denotes the (formal) covariant 
metric derivative with respect to $g$. As it is usual in General 
Relativity, all expressions in terms of $D_{\mu}$ have an equivalent 
counterpart in terms of $\nabla_{\mu}$.

A ``horizontal'' superpotential is given by:
\bEq
\eta_{h}^{\sig\mu}=\frac{\sqrt{g}}{4\kappa}
(\nabla^{\sig}\xi^{\mu}-\nabla^{\mu}\xi^{\sig})\,,
\eEq
which is essentially the Komar superpotential \cite{Kom59}.

Furthermore, we have
\bEq
\eps^{\sig}(\lam_{YM},\Xi_{v})=
-2p^{\mu\sig}_{A}\nabla_{\mu}\xi^{A}_{v}= 
-\nabla_{\mu}(-2p^{\mu\sig}_{A}\xi^{A}_{v})+ 
2\nabla_{\mu}p^{\mu\sig}_{A}\xi^{A}_{v} \,.
\eEq
Then there exists a ``vertical'' superpotential, given by:
\beq
\eta_{v}^{\mu\nu}=p^{\mu\nu}_{A}\xi^{A}_{v}=-\frac{\sqrt{g}}{2}F^{\mu\nu}_{A}
\xi^{A}_{v}\,.
\eeq

\bRm
Notice that here the splitting of the current and of the superpotential 
is due to the fact that the bundle $\bY$ is split from the beginning.
\END
\eRm 

\medskip

{\em Acknowledgments.}
Thanks are due to M. Ferraris, I. Kol\'a\v r and D. Krupka for 
interesting discussions. 
Special thanks are due to R. Vitolo for helpful comments. The 
authors would also like to thank the referee for useful suggestions.



\footnotesize  

\begin{thebibliography}{}

\bibitem{Ec81}{\sc D.J. Eck}: Gauge--natural bundles and generalized 
gauge theories, {\em Mem. Amer. Math. Soc.} {\bf 247} (1981)
1--48.

\bibitem{Fa99} {\sc L. Fatibene}: Formalismo gauge--naturale per 
le teorie di campo classiche. Ph.D. Thesis, University of Torino (1999).

\bibitem{Fe84}{\sc M. Ferraris}: Fibered Connections and Global
Poincar\'e--Cartan Forms in Higher--Order Calculus of Variations, in: 
{\em Proc. Conf. Diff. Geom. and Appl.} (Nov\'e M\v{e}sto na 
Morav\v{e}, 1983); D. Krupka ed.; J. E. Purkyn\v{e} University (Brno, 
1984) 61--91.

\bibitem{FeFr91}{\sc M. Ferraris, M. Francaviglia}: The Lagrangian 
Approach to Conserved Quantities in General Relativity, in:
{\em Mechanics, Analysis and Geometry: 200 Years after Lagrange}; M. 
Francaviglia ed.; Elsevier Science Publishers B. V. (Amsterdam, 1991), 
451--488.

\bibitem{FFR83}{\sc M. Ferraris, M. Francaviglia, C. Reina}: Sur les 
fibr\'es d'objects g\'eom\'etriques et leurs applications 
physiques, {\em Ann. Inst. Henri Poincar\'e} {\bf 38} (4) (1983) 
371--383.

\bibitem{FFR84}{\sc M. Ferraris, M. Francaviglia, O. Robutti}: Energy 
and Superpotentials in Gravitational Theories, in: {\em Atti del VI 
convegno nazionale di Relativit\`a Generale  e Fisica della 
Gravitazione},
(Firenze, 1984); M. Modugno ed.; Pitagora Editrice (Bologna, 1986) 
137--150.

\bibitem{FPV98a}{\sc M. Francaviglia, M. Palese, R. Vitolo}: 
Symmetries in finite order variational sequences, to appear in 
{\em Czech. Math. Journ.}.

\bibitem{FPV98b}{\sc M. Francaviglia, M. Palese, R. Vitolo}: 
Superpotentials in variational sequences, {\em Proc. 
VII Conf. Diff. Geom. and Appl., Satellite Conf. of ICM in Berlin} 
(Brno 1998); I. Kol\'a\v r {\em et al.} eds.; Masaryk University in 
Brno (Czech Republic) 1999, 469--480.

\bibitem{GMV91} {\sc G. Giachetta, L. Mangiarotti, R. Vitolo}: 
The Einstein--Yang--Mills Equations, {\em
Gen. Rel. and Grav.} {\bf 23} (6) (1991) 641--659.

\bibitem{HoKo83}{\sc M. Hor{\'a}k, I. Kol{\'a}{\v r}}: On the Higher 
Order Poincar\'e--Cartan Forms, {\em Czech. Math. Journ.}, 
{\bf 33} (108) (1983) 467--475.

\bibitem{Ja90} {\sc J. Jany\v{s}ka}: Natural and Gauge--Natural Operators 
on the Space of Linear Connections on a Vector Bundle, 
{\em Proc. Diff. Geom. and its Appl.} (Brno, 1989); 
J. Jany\v{s}ka, D. Krupka eds.; World Scientific (Singapore, 1990) 
58--68.

\bibitem{KoNo63} {\sc S. Kobayashi, K. Nomizu}: 
{\em Foundations of Differential Geometry} vol. {\bf I}, Interscience
Publishers (John Wiley \& Sons, 1963).

\bibitem{KMS93} {\sc I. Kol{\'a}{\v r}, P.W. Michor, J. Slov{\'a}k}:
{\em Natural Operations in Differential Geometry},
(Springer--Verlag, N.Y., 1993).

\bibitem{Kol75}{\sc I. Kol\' a\v{r}}: On some operations with 
connections, {\em Math. Nachr.}, {\bf 69} (1975) 297--306.

\bibitem{Kol79} {\sc I. Kol\' a\v{r}}: Prolongations of generalized 
connections, {\em Coll. Math. Soc. J\'anos Bolyai}, 
(Differential Geometry, Budapest, 1979) 
{\bf 31} (1979) 317--325.

\bibitem{Kol83}{\sc I. Kol\' a\v{r}}: A Geometrical Version of the 
Higher Order Hamilton Formalism in Fibred Manifolds, {\em J. Geom. 
Phys.}, {\bf 1} (2) (1984) 127--137.

\bibitem{Kol84}{\sc I. Kol\' a\v{r}}: Some Geometric Aspects of
the Higher Order Variational Calculus, {\em Geom. Meth. in Phys., 
Proc. Diff. Geom. and its Appl.}, (Nov\'e M\v{e}sto na 
Morav\v{e}, 1983); D. Krupka ed.; J. E. Purkyn\v{e} University (Brno, 
1984) 155--166.

\bibitem{Kom59}{\sc A. Komar}:
Covariant Conservation Laws in General Relativity,
{\em Phys. Rev.} {\bf 113} (1959) (3) 934--936.

\bibitem{Kru74}{\sc D. Krupka}: A Setting for Generally Invariant 
Lagrangian Structures in Tensor Bundles, {\em Bull. Acad. Pol. Sc., 
Ser. Sc. Math., Astr. et Phys.} {\bf XXII} (9) (1974) 967--972.
 
\bibitem{Kru90}{\sc D. Krupka}: Variational Sequences on Finite
Order Jet Spaces, {\em Proc. Diff. Geom. and its
Appl.} (Brno, 1989); J. Jany\v{s}ka, D. Krupka eds.; World Scientific
(Singapore, 1990) 236--254.

\bibitem{Kru93}{\sc D. Krupka}: Topics in the Calculus of
Variations: Finite Order Variational Sequences, {\em Proc. Diff. 
Geom. and its Appl.} (Opava, 1993) 473--495.

\bibitem{KrJa90}{\sc D. Krupka, J. Jany\v{s}ka}: Lectures on 
Differential Invariants, Univerzita J. E. Purkyn\v{e} V Brn\v{e} 
(1990).

\bibitem{MaMo83a}{\sc L. Mangiarotti, M. Modugno}:
Fibered Spaces, Jet Spaces and Connections for Field Theories, in
{\em Proc. Int. Meet. on Geom. and Phys.}; M. Modugno ed.; Pitagora Editrice
(Bologna, 1983) 135--165.

\bibitem{Pa00} {\sc M. Palese}: Geometric Foundations of the Calculus 
of Variations. Variational Sequences, Symmetries and Jacobi Morphisms.
Ph.D. Thesis, University of Torino (2000).

\bibitem{Sau89}{\sc D.J. Saunders}: The Geometry of Jet Bundles,
Cambridge Univ. Press (Cambridge, 1989).

\bibitem{Vit97}{\sc R. Vitolo}: On Different Geometric Formulations 
of Lagrangian Formalism, {\em Diff. Geom. and its Appl.} {\bf 10} (1999) 
225--255.

\bibitem{Vit98}{\sc R. Vitolo}: Finite Order Lagrangian Bicomplexes,
{\em Math. Proc. Camb. Phil. Soc.} {\bf 125} (1) (1999) 321--333.

\bibitem{Wel80}{\sc R O. Wells}: Differential Analysis on Complex
Manifolds, {\em GTM} n. {\bf 65}, Springer--Verlag (Berlin, 1980).

\end{thebibliography}
\end{document}